%% file: ams01isotopes.tex
\documentclass[apj]{emulateapj}           
\slugcomment{Draft version \-- NT May 25$^{\rm{th}}$ 2011 }
\usepackage{graphicx,amssymb,amsmath,times}
\usepackage{epsfig,epstopdf}
\usepackage[colorlinks=true,pdfstartview=FitV,linkcolor=blue,citecolor=blue,urlcolor=blue]{hyperref}  

\def\Journal#1#2#3#4{{#4}, {#1}, {#2}, #3} 

\newcommand{\etal}{et al.}
\newcommand{\mysize}{\footnotesize}

\begin{document}

\title{Isotopic Composition of Light Nuclei  
in Cosmic Rays: Results from AMS-01}         
\shorttitle{Light Isotopes in Cosmic Rays}   
\shortauthors{Aguilar \etal}                 
\input authorlist_eapj                       

\begin{abstract} 

The variety of isotopes 
in cosmic rays allows us to study different aspects of the processes that 
cosmic rays undergo between the time they are produced and the time of their arrival in the heliosphere.
In this paper we present measurements of the isotopic ratios
$^{2}$H/$^{4}$He, $^{3}$He/$^{4}$He, $^{6}$Li/$^{7}$Li, $^{7}$Be/($^{9}$Be+$^{10}$Be) and $^{10}$B/$^{11}$B 
in the range $0.2\--1.4$ GeV of kinetic energy per nucleon. 
The measurements are based on the data collected by the 
Alpha Magnetic Spectrometer, AMS-01, during the STS-91 flight in 1998 June.

\end{abstract}

\keywords{acceleration of particles --- cosmic rays --- nuclear reactions, nucleosynthesis, abundances}

\section{Introduction}    
\label{Sec::Introduction} 

Cosmic rays (CRs) detected with kinetic energies in the range from MeV to TeV per nucleon
are believed to be produced by galactic sources.
Observations of X-ray and $\gamma$-ray emission from galactic sites
such as supernova remnants, pulsars or stellar winds
reveal the presence of energetic particle acceleration 
mechanisms occurring in such objects.
The subsequent destruction of these accelerated nuclei (\textit{e.g.,} p, He, C, N, O, Fe) in the
interstellar medium gives rise to secondary species
that are rare in the cosmic ray sources, such as Li, Be, B, sub-Fe elements, 
deuterons, antiprotons, positrons and high energy photons.
The relation between secondary CRs and their primary progenitors
allows the determination of propagation parameters, such as the 
diffusion coefficient and the size of the diffusion region.
For a recent review, see \citet{Strong2007}.

Along with the ratios B/C and sub-Fe/Fe,
it is of great importance to determine the propagation 
history of the lighter H, He, Li and Be isotopes.
Since $^{2}$H and $^{3}$He CRs are mainly produced by the breakup of the primary $^{4}$He in the galaxy,
the ratios $^{2}$H/$^{4}$He and $^{3}$He/$^{4}$He probe the propagation history of helium~\citep{Webber1997}.
The isotopes of Li, Be and B, all of secondary origin, 
are also useful for a quantitative understanding of CR propagation. 
The relative abundances and isotopic composition of H, He, Li, Be and B, therefore,
might help to distinguish between the propagation models and give 
constraints to their parameters \citep{Moskalenko2003ICRC}. 

Low energy data ($\lesssim$200 MeV nucleon$^{-1}$) on CR isotopic composition come mainly from space experiments 
such as the HET telescopes on VOYAGER 1 and 2 \citep{Webber2002}, the Cosmic Ray Isotope Spectrometer (CRIS) on the 
Advanced Composition Explorer (ACE) satellite~\citep{DeNolfo2001}, 
the ULYSSES high energy telescope~\citep{Connell1998}
and the HKH experiment on the International Sun-Earth Explorer (ISEE) spacecraft~\citep{Wiedenbeck1980}.
Light nuclei data at higher energies (up to few GeV~nucleon$^{-1}$) 
have been measured by balloon borne magnetic spectrometers
including IMAX~\citep{Reimer1998}, ISOMAX~\citep{Hams2004}, SMILI~\citep{Ahlen2000}, BESS~\citep{Wang2002}, 
Inteplanetary Monitoring Platform (IMP) experiment~\citep{GarciaMunoz1977} and the
Goddard Space Flight Center (GSFC) balloon~\citep{Hagen1977}.

AMS-01 observed CRs at an altitude of $\sim\,$380 km during a period, 1998 June,
of relatively quiet solar activity.
It collected data free from atmospheric induced background.
In this paper
we present measurements of the $^{3}$He/$^{4}$He ratio over the
kinetic energy range $0.2\--1.4$ GeV per nucleon, and the average values of the 
ratios $^{6}$Li/$^{7}$Li, $^{7}$Be/($^{9}$Be+$^{10}$Be) and $^{10}$B/$^{11}$B 
over the same energy range. The ratio $^{2}$H/$^{4}$He is also presented.

\section{The Alpha Magnetic Spectrometer} 
\label{Sec::AMS}                          

The Alpha Magnetic Spectrometer (AMS) is a particle physics instrument designed for the 
high precision and long duration measurement of CRs in space.
The AMS-01 precursor experiment operated successfully during a 10 day flight on
the space shuttle \textit{Discovery} (STS-91). 

The spectrometer was composed of a cylindrical permanent magnet, a silicon micro-strip tracker,
time-of-flight (TOF) scintillator planes, an aerogel \v{C}erenkov counter and anti-coincidence counters.
The performance of AMS-01 is described elsewhere~\citep{AMS01Report2002}.

Data collection started on 1998 June 3. The orbital
inclination was 51$^{\circ}$.7 and the geodetic altitude ranged
from 320 to 390 km. The data were
collected in four phases: 
(a) 1 day of check out before docking with the MIR space station, 
(b) 4 days while docked to MIR, 
(c) 3.5 days with AMS pointing directions within 0$^{\circ}$, 20$^{\circ}$ and 45$^{\circ}$ of the zenith 
and (d) 0.5 days before descending, pointing toward the nadir.

The acceptance criterion of the trigger logic in the AMS-01 instrument was
a four-fold coincidence between the signals from the four TOF planes. 
Only particles traversing the silicon tracker were accepted.
Events crossing the anti-coincidence counters or producing multiple hits in the TOF  
layers were rejected.
A prescaled subsample of 1 out of 1000 events was recorded with a dedicated minimum-bias 
configuration. This ``unbiased trigger'' required only the TOF coincidence.

The AMS-01 mission provided results on cosmic ray protons, 
helium, electrons, positrons and light nuclei~\citep{AMS01Report2002,AMS01Positron2007,AMS01Nuclei2010}. 
During the flight, a total of 99 million triggers were recorded by the spectrometer,
with 2.85M helium nuclei and nearly 200,000 nuclei with charge $Z>2$. 

\section{Data Analysis}   
\label{Sec::DataAnalysis} 

The identification of cosmic ray nuclei with AMS-01 was performed
through the combination of independent measurements provided by the various detectors.
The particle rigidity, $R$, (momentum per unit charge, $pc/Ze$) was provided by the deflection of the
reconstructed particle trajectory in the magnetic field.
The velocity, $\beta=v/c$, was measured from the particle transit time between the four TOF
planes along the track length. 
The reconstruction algorithm provided, together with the
measured quantities $R$ and $\beta$, an estimation of their uncertainties 
$\delta R$ (from tracking) and $\delta \beta$ (from timing), 
that reflected the quality of the spectrometer in performing such measurements.
The particle charge magnitude $|Z|$ was obtained by the analysis of the multiple 
measurements of energy deposition in the four TOF scintillators up to $Z=2$~\citep{AMS01Report2002}
and the six silicon layers up to $Z=8$~\citep{AMS01Nuclei2010}.
The particle mass number, $A$, was therefore determined from the resulting charge, velocity and rigidity: 
\begin{equation}
  A = \frac{R Z e}{m_{n}\beta c^{2}}\sqrt{ 1 - \beta^{2} } 
  \label{Eq::Mass}
\end{equation} 
where $m_{n}$ is the nucleon mass.

The response of the detector was simulated using the AMS simulation program, 
based on \texttt{GEANT-3.21}~\citep{GEANT3} 
and interfaced with the hadronic package \texttt{RQMD} (relativistic quantum molecular dynamics~\citep{RQMD}). 
The effects of energy loss, multiple scattering, nuclear interactions and decays 
were included, as well as detector efficiency and resolution.
After the flight, the detector was extensively calibrated 
at GSI, Darmstadt, with ion beams (He, C) and at the CERN-PS, Geneva,
with proton beams. This ensured that the performance of the detector and 
the analysis procedure were thoroughly understood.

Further details are found in \citet{AMS01Report2002} and references therein.

\subsection{Helium Isotopes} 
\label{Sec::HeliumIsotopes}  

Given the large amount of statistics available for $Z=2$ data,  
we considered only the highest quality data collected during the post-docking phase 
(c) and only while pointing toward the zenith.
Data taken while passing near the South Atlantic Anomaly 
(latitude: $5^{\circ}\--45^{\circ}\,$S, longitude: $5^{\circ}\--85^{\circ}\,$W) were excluded.
Only events taken when the energy interval $0.2\--1.4$ GeV~nucleon$^{-1}$ was above the geomagnetic 
cutoff for both the isotopes $^{3}$He and $^{4}$He were kept;  
this corresponds in selecting the orbital regions with the highest geomagnetic latitudes, 
$\Theta_{\textrm{M}}$, roughly $|\Theta_{\textrm{M}}| \gtrsim 0.9$.

Furthermore, the acceptance was restricted to particles traversing the detector top-down 
within $30^{\circ}$ of the positive $z$-axis.
Events with poorly reconstructed trajectories were rejected through quality cuts on the associated $\chi^{2}$
or consistency requirements between the two reconstructed half tracks~\citep{AMS01Nuclei2010}. 
To avoid biasing the reconstructed mass distributions, 
no cuts on the consistency of the TOF velocity versus tracker rigidity measurements were applied. 
We required that the velocity was measured with hits from at least three out of four TOF planes 
and that the
rigidity was reconstructed with at least five out of six tracker layers. 
Approximately 18,000 nuclei with charge $Z=2$ were selected in the energy range $0.2\--1.4$ GeV~nucleon$^{-1}$.
The charge was determined from the energy depositions in both the TOF and tracker layers.
The kinetic energy per nucleon was measured with the TOF system, \textit{i.e.,} through the velocity $\beta$.
In the considered energy range, the TOF energy resolution is comparable to that of the tracker.

\begin{figure}[!h]
\begin{center}
\epsscale{0.950}
\plotone{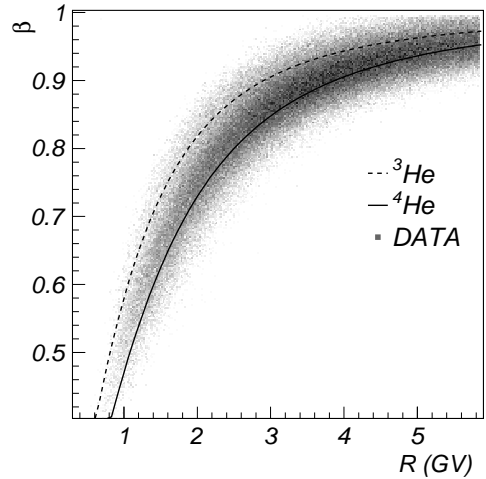}
\figcaption{ 
  Distribution of the measured velocity, $\beta$, as a function of the reconstructed rigidity, $R$, for $Z=2$ nuclei.
  The two lines represent the exact relationship of $\beta$ to $R$ for the two isotopes 
  $^{3}$He (dashed line) and $^{4}$He (solid line).
  \label{Fig::BetaVSRig}
}
\end{center}
\end{figure}

The selected data are shown in Fig.~\ref{Fig::BetaVSRig} distributed in the $(\beta,R)$ plane.
The two curves represent the exact relation between velocity $\beta$ and rigidity $R$ for a $Z=2$ 
nucleus of mass number $A=3$ (dashed line) and $A=4$ (solid line), which is:
\begin{equation}
  \beta = \left[ 1 + A^{2}\left( \frac{m_{n}c^{2}}{Ze R} \right)^{2} \right]^{-1/2}
\label{Eq::BetaVSRig}\end{equation}
The large dispersion of the measured data, apparent from Fig.~\ref{Fig::BetaVSRig}, indicates 
a relatively poor mass resolution in the separation of the two mass numbers.
Under these conditions, 
any event-to-event separation (\textit{e.g.,} through a mass cut) is clearly inapplicable.
In addition, the distribution of the reconstructed mass numbers (Eq.~\ref{Eq::Mass}) exhibits asymmetric tails; 
so the standard Gaussian fit method~\citep{Seo1997} is not appropriate
for describing the observed mass response of the instrument.

\begin{figure}[!h]
\begin{center}
\epsscale{0.950}
\plotone{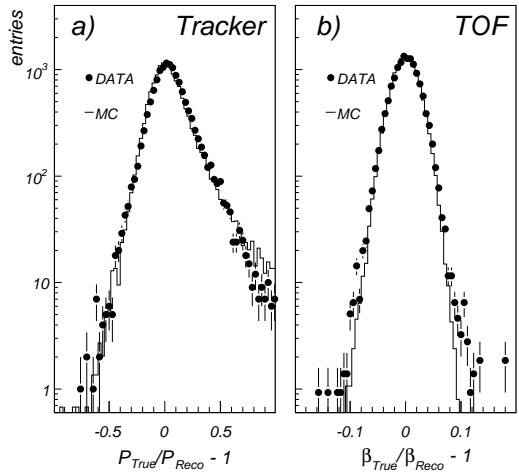}
\figcaption{
  a) Rigidity and b) velocity resolutions of the AMS-01 tracker and TOF estimated with 
  measured data from the test beam with $E=2$ GeV~nucleon$^{-1}$ helium nuclei.
  Data (circles) are compared with the MC simulation (histograms). 
  The MC entries are normalized to the data.
  \label{Fig::ccResolutionHelium}
}
\end{center}
\end{figure}

In order to determine the isotopic ratios, it was therefore necessary 
to develop a comprehensive model for the complete response of the instrument to different masses.
The mass resolution is influenced by the intrinsic time resolution of the TOF system and
by the bending power of the magnet coupled with the intrinsic spatial resolution of the tracker.
Physical processes such as multiple scattering, energy losses and interactions
along the particle path also contribute in shaping the reconstructed mass distributions.
Thus, we modeled the AMS-01 mass response by means of our Monte Carlo (MC) simulation program,
which includes all the aforementioned physical effects as well as the instrumental readout, 
providing a realistic description of particle tracking and timing on an event-by-event basis.
The program was also tuned with data collected during the test beams. 
The resulting rigidity resolution $\delta R/R$ and velocity resolution $\delta \beta / \beta$ are 
shown in Fig.~\ref{Fig::ccResolutionHelium} for test beam data (filled circles) and MC events (histograms). 
It can be seen that the mass resolution, approximately given by:
\begin{equation}
\left( \frac{\delta A}{A} \right)^{2} = \left( \gamma^{2} \frac{\delta \beta}{\beta} \right)^{2} + \left( \frac{\delta R}{R} \right)^{2} \; , 
\end{equation}
was correctly simulated as the MC agrees with the data within $\sim\,$2\%.

\begin{figure*}[!ht]
\begin{center}
\epsscale{0.900}
\plotone{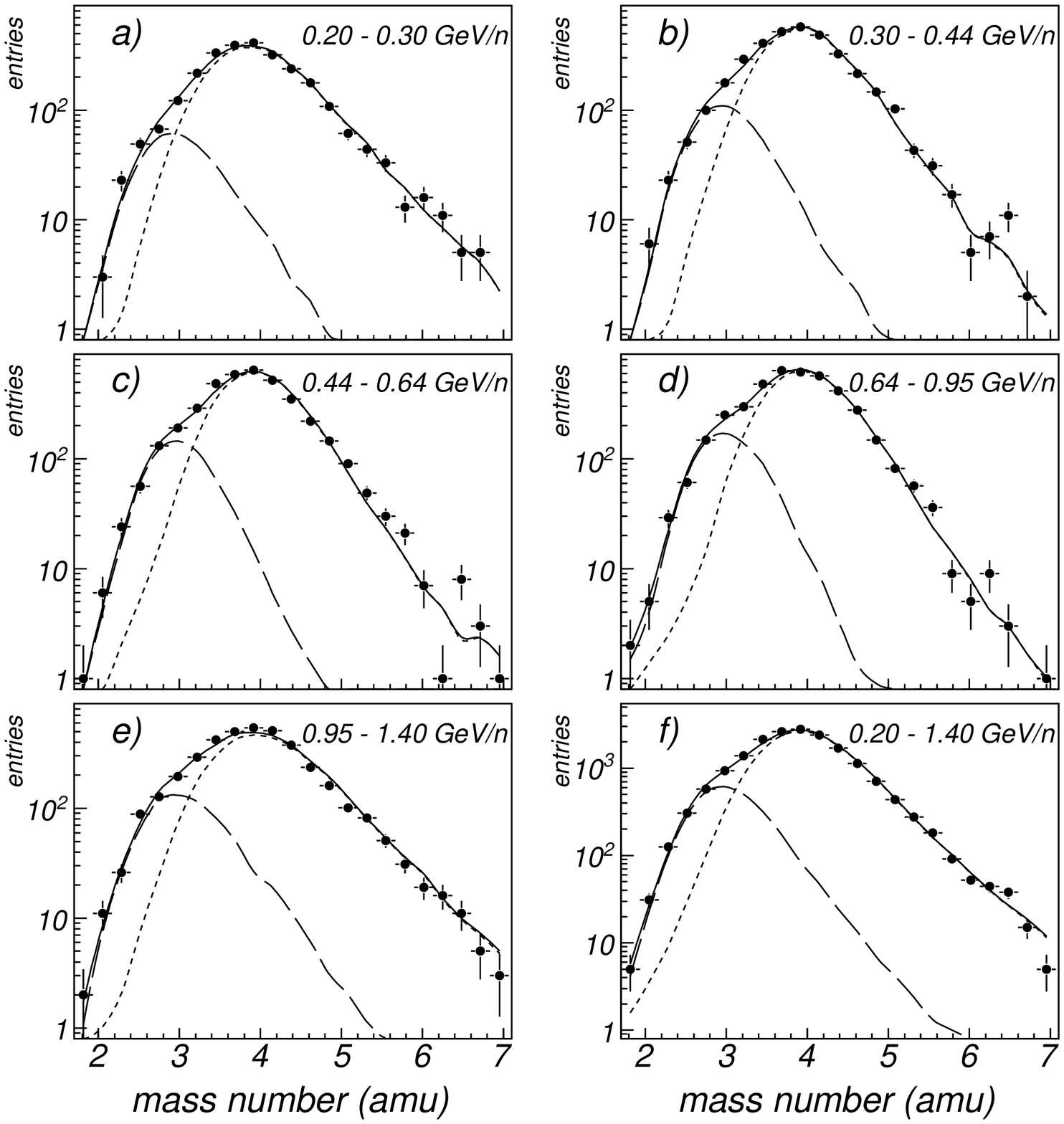}
\figcaption{ 
  Reconstructed mass distributions of $Z_{rec}=2$ events from flight data (solid circles)
  and from MC generated isotopes of $^{3}$He (long-dashed lines), $^{4}$He (short-dashed lines) 
  and their sum (solid lines).
  Distributions are shown in five energy intervals from $a)$ to $e)$ and over the entire range $f)$
  between 0.2 and 1.4 GeV~nucleon$^{-1}$.
  \label{Fig::HeliumMasses}
}
\end{center}
\end{figure*}

Using a sufficiently large number of simulated events of $^{3}$He and $^{4}$He
and with the ratio $^{3}$He/$^{4}$He of the detected events as a free parameter, 
we determined the best composition fit between the simulated mass 
distributions and the measured one.
In these fits the overall normalization, $\mathcal{N}$, was also a free parameter.
In principle, $\mathcal{N}$ should be fixed by the data,
namely by the number of entries, $\mathcal{N}_{\textrm{E}}$, of each mass histogram. 
Deviations of $\mathcal{N}$ from its expected value $\mathcal{N}_{\textrm{E}}$ may indicate the 
presence of an unaccounted background, \textit{e.g.,} from charge misidentification.

The results of this procedure are shown in Fig.~\ref{Fig::HeliumMasses}, where the agreement between the 
measured mass histograms (filled circles) and the simulated histograms (lines) 
turned out to be very satisfactory.
The fits of the $^{3}$He/$^{4}$He mass composition ratios gave  
unique minima in all the considered energy bins.
The uncertainties associated to these ratios were directly determined
from the 1-$\sigma$ uncertainties in the $\chi^{2}$ statistics of
the fitting procedure. 
The $\chi^{2}$ fitting method was cross checked with the Maximum Likelihood 
method. The two methods gave the same results and very similar uncertainties. 
Double Gaussian fits were also performed in order to provide the
corresponding mass resolution, $\delta$A/A, for each energy bin,
defined as the ratio between the width and the mean\footnote[1]{
  Equal mass resolutions were obtained for $^{3}$He and $^{4}$He within 0.1\% at all energies. Table~\ref{Tab::He3He4RatioSummary} provides the mean values.
}.
The fitted $^{3}$He/$^{4}$He ratios for all the considered energy bins
from 200 MeV nucleon$^{-1}$ to 1.4 GeV~nucleon$^{-1}$ are listed in Table~\ref{Tab::He3He4RatioSummary},
together with the $\chi^{2}$/df values, the number of events,
the mass resolution and correction factors discussed below.

\setlength{\tabcolsep}{0.04in} 
\begin{deluxetable}{ccccccc}[!h]
\tablecaption{
  Fit results and correction factors for the ratio $^{3}$He/$^{4}$He
  between 0.2 and 1.4 GeV of kinetic energy per nucleon.
\label{Tab::He3He4RatioSummary}}
\tablehead{
  \colhead{Energy} & \colhead{Events} & \colhead{Fit Results} & \colhead{$\chi^{2}$/df}  & 
  \colhead{$\delta$A/A} & \colhead{ACorr} & \colhead{FCorr}
}
\startdata
  0.20--0.30 & 2,660 & 0.125 $\pm$ 0.011 & 32.3/29 &13.1\% & 1.12 & 0.97 \\
  0.30--0.44 & 3,553 & 0.158 $\pm$ 0.096 & 51.1/29 &12.2\% & 1.05 & 0.98 \\
  0.44--0.64 & 3,867 & 0.182 $\pm$ 0.094 & 65.8/34 &11.8\% & 1.00 & 0.98 \\
  0.64--0.95 & 4,142 & 0.211 $\pm$ 0.098 & 62.7/30 &12.2\% & 0.99 & 0.98 \\
  0.95--1.40 & 3,813 & 0.223 $\pm$ 0.012 & 55.0/33 &13.9\% & 0.99 & 0.98 
\enddata
\end{deluxetable}

\subsection{Top-Of-Instrument Corrections} 
\label{Sec::TOICorrections}                

The measured mass distribution of Fig.~\ref{Fig::HeliumMasses} was fitted with
an MC sample of mixed $^{3}$He and $^{4}$He that
were sent through the same analysis chain (trigger, reconstruction and data selection) as the data.
The free parameter is the ratio $^{3}$He/$^{4}$He of the two mass distributions corresponding to
the recorded events. 
Hence, small corrections have to be performed in order to extract the ratio of interest, namely the ratio entering the instrument. 
These are referred to as Top-Of-Instrument (TOI) corrections. 

\subsubsection{Acceptance Corrections} 
\label{Sec::Acceptance}                

The detector acceptance, $\mathcal{A}$, includes trigger efficiency, 
reconstruction efficiency and selection efficiency.
The acceptance was calculated using our Monte Carlo simulation program. 
Nucleus trajectories were simulated 
in the energy range $\sim 0.05 \-- 40$~GeV~nucleon$^{-1}$.
They were emitted downward from a square of length 3.9 m placed above the detector. 
In total, 110 million $^{3}$He and 550 million $^{4}$He nuclei were simulated. 
The physical processes involved and the detector response are very similar for the two isotopes, 
\textit{i.e.,} the resulting acceptances are quite similar in magnitude. 
The contributions from the detector
acceptance mostly cancel in the ratio.  The associated systematic errors also cancel.
The only important factor in determining the isotopic ratios is the knowledge of any
isotopic dependent effects in the detector response.
Mass dependent features are expected from the following effects:

\begin{itemize}
\item \textit{Rigidity threshold}.
The instrument acceptance is rigidity dependent, because the tracks of slower particles are more curved,
and it is less likely that they pass through both the upper and lower TOF counters and the tracking volume.
Since, at the same kinetic energy per nucleon, the lighter isotope $^{3}$He 
have lower rigidity than the heavier $^{4}$He, 
the resulting acceptance, particularly at lower energies, is lower for $^{3}$He. 
Above $0.2$~GeV~nucleon$^{-1}$, the rigidity threshold affects the ratio by less than $\sim\,$1\%. 

\item \textit{Multiple scattering}.
  Coulomb scattering is slightly more pronounced for lighter particles. 
  Since multiple scattering affects the event reconstruction and selection efficiency, 
  the acceptance for the lighter isotope is smaller in the lowest energy region.
  The multiple scattering effect amounts to $\sim\,$10\% at E$\sim\,$0.2 GeV/n
  and decreases with energy, down to $\sim\,$1\% at E$\sim\,$1 GeV/n.
\item \textit{Nuclear interactions}.
  The attenuation of cosmic rays after traversing the TOI material
  is isotope dependent and closely related to the inelastic cross
  sections, $\sigma_{\textrm{\tiny int}}$, for the interactions in the
  various layers of the detector material.
  For $^{3}$He and $^{4}$He, the attenuation due to interactions differs by $\sim\,$2\% or less.
\end{itemize}
\begin{figure}[!ht]
  \begin{center}
    \epsscale{0.950}
    \plotone{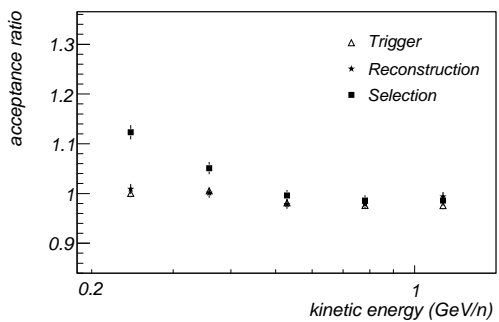}
    \figcaption{ 
      Ratios of the acceptances of $^{4}$He and $^{3}$He as a function of kinetic energy per nucleon. 
      This quantity is shown after the successive application of the trigger (open triangles), reconstruction (stars), and 
      selection (filled squares) cuts.
      \label{Fig::HeliumAccRatios}
    }
  \end{center}
\end{figure}
Fig.~\ref{Fig::HeliumAccRatios} shows the ratio of the two acceptances
for MC events
after the successive application of the trigger, reconstruction and then selection cuts.
Deviations are appreciable below 0.4 GeV~nucleon$^{-1}$ and mainly due to the event selection (filled squares).
This indicates the dominance of the ``multiple scattering effect'' mentioned above, 
because the selection cuts acted against events with large scattering angles.

These mass dependent features, due to the particle dynamics in 
the detector, did not appreciably influence the AMS-01 trigger system.
\begin{figure}[!h]
\begin{center}
\epsscale{0.950}
\plotone{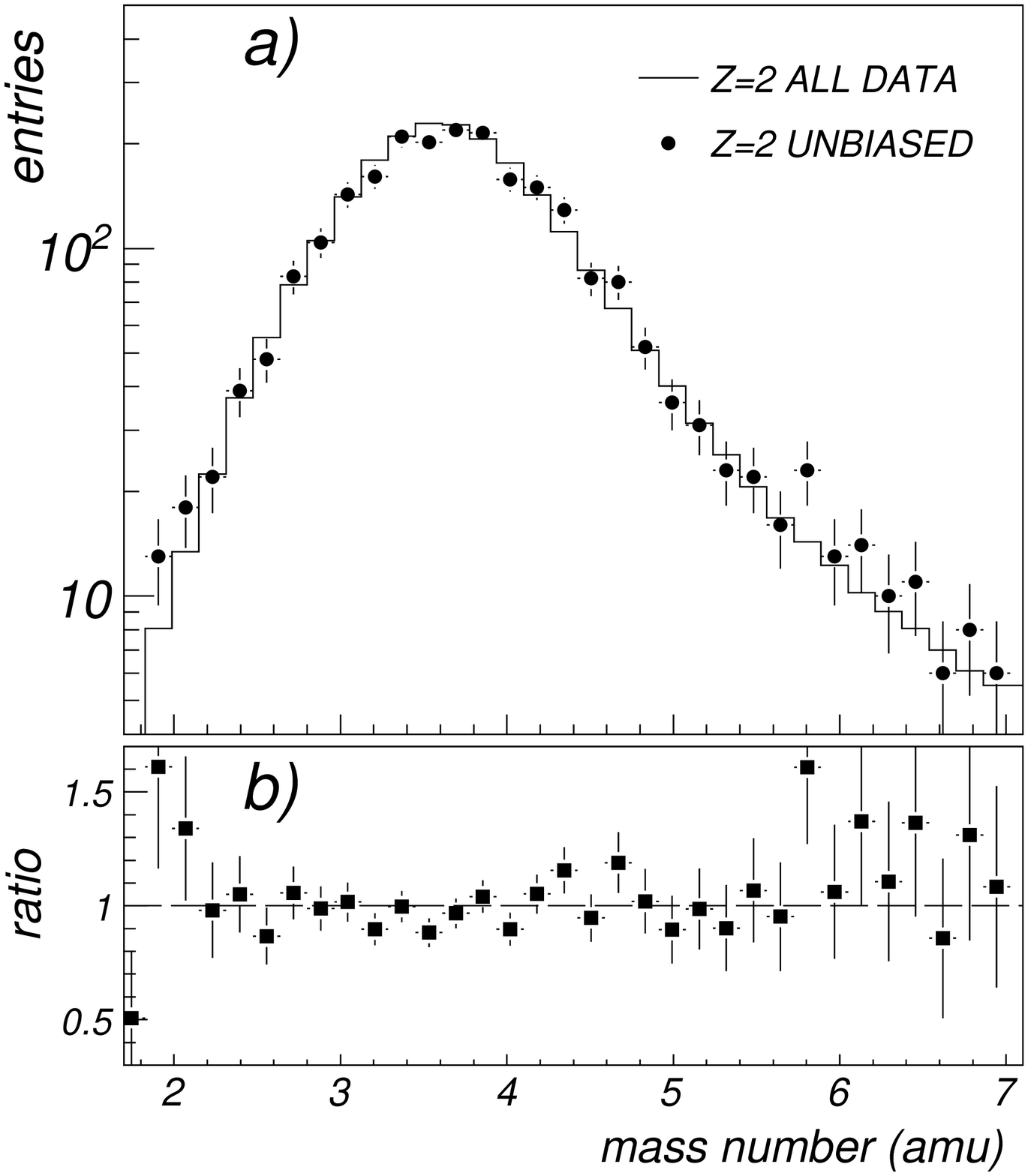}
\figcaption{ 
  a) Reconstructed mass distribution of $Z=2$ events collected with the unbiased trigger (filled circles) 
  in comparison with the $Z=2$ mass histogram obtained with all the data (histogram). 
  b) Ratio of the two histograms (squares); 
  the horizontal dashed line is shown for reference.
  \label{Fig::ccUnbiasedHe}
}
\end{center}
\end{figure}
In Fig.~\ref{Fig::ccUnbiasedHe} we report the reconstructed mass distribution 
using data collected with the unbiased trigger (\S\ref{Sec::AMS}). 
The comparison of such unbiased data (circles) with all $Z=2$ data collected 
from the flight (solid line, normalized to the unbiased data entries) shows
no significant difference in the mass distribution.

\subsubsection{Nuclear Interactions} 
\label{Sec::NuclearInteractions}     

Table~\ref{Tab::Material} lists the material between the top of the payload and the tracker:
a Multi-Layer Insulation (MLI) blanket,
a Low Energy Particle (LEP) shield 
and two TOF layers of plastic scintillators supported by a honeycomb structure.   
In total there were $\sim\,$5 g/cm$^{2}$ of material above the tracking volume. 
\begin{deluxetable}{lll}[!h]
\tablecaption{
  Material above the tracker.
  The column density is averaged over the angle of incidence. 
  \label{Tab::Material}}
\tablehead{
  \colhead{Detector Element}  &  \colhead{Composition}  &  \colhead{Amount} 
}
\startdata
MLI thermal blanket    &    C$_5$ H$_4$ O$_2$ &  0.7 $g/cm^2$   \\
LEP shield             &    C                &  1.3 $g/cm^2$   \\
TOF scintillator       &    C/H$ = 1$        &  2.1 $g/cm^2$   \\
TOF support structure  &    Al               &  1.0 $g/cm^2$   
\enddata
\end{deluxetable}
The \texttt{RQMD} interface used in the AMS simulation program provided a
simulation of all the high energy hadronic collisions involving deuterons, $^{3}$He, $^{4}$He and heavier ions.
These effects give an appreciable contribution to the total acceptance of \S\ref{Sec::Acceptance}. 
The survival probability of $^{3}$He ($^{4}$He) after traversing the TOI material of Table~\ref{Tab::Material}
varies between $\sim\,$90\% ($\sim\,$88\%) at $\sim\,$0.2~GeV~nucleon$^{-1}$ and
$\sim\,$86\% ($\sim\,$86\%) at $\sim\,$1.4~GeV~nucleon$^{-1}$.

More critical is the fragmentation of $^{4}$He into $^{3}$He, that required a dedicated correction. 
In this process, if only a neutron is ``stripped'' above the tracker, 
the event is recorded as a clean $^{3}$He event. 
The measured $^{3}$He/$^{4}$He ratio is then distorted by incoming $^{4}$He that spill over into the $^{3}$He mass distribution.
Note that the simulated mass distributions of Fig.~\ref{Fig::HeliumMasses} are referred to the particle identities 
within the tracking volume, \textit{i.e.,} the $^{3}$He mass histograms of the figures (long-dashed lines) also contain the 
``extra'' $^{3}$He nuclei generated as fragmentation products of $^{4}$He. 
Assuming that the kinetic energy per nucleon is maintained in the mass changing process, 
the ratio has been corrected for this effect.
For each considered energy interval, the ratio $\eta$ between the ``extra'' $^{3}$He  
and the total number of detected $^{4}$He was estimated.
The isotopic ratio $^{3}$He/$^{4}$He resulting from the composition fit $\mathcal{M}$ is then related to the 
TOI ratio $\mathcal{R}$ through the relation:
\begin{equation}
\mathcal{M} = \frac{ \mathcal{A}_{3} \,\phi_{3} + \eta \mathcal{A}_{4}  \, \phi_{4} }{ \mathcal{A}_{4} \, \phi_{4} -  \eta \mathcal{A}_{4} \, \phi_{4} } = 
\left( \frac{1}{1 - \eta} \right) \left[ \left( \frac{\mathcal{A}_{3}}{\mathcal{A}_{4}} \right) \mathcal{R}  + \eta \right] 
\label{Eq::MeasuredVSTOI}
\end{equation} 
where $\phi_{3}$ and $\phi_{4}$ are the incident (TOI) intensities of the two isotopes ($\mathcal{R}\equiv\phi_{3}/\phi_{4}$) and 
$\mathcal{A}_{3}$ and $\mathcal{A}_{4}$ the corresponding acceptances.  
Inverting Eq.~\ref{Eq::MeasuredVSTOI} we obtain the TOI ratio:
\begin{equation}
\mathcal{R}= \left( \frac{\mathcal{A}_{4}}{\mathcal{A}_{3}} \right) \left[ 1 - \eta - \frac{\eta}{\mathcal{M}} \right] \mathcal{M}
\label{Eq::TOIVSMeasured}
\end{equation}
We then define the TOI correction factors as 
\texttt{ACorr}~$\equiv$~$\mathcal{A}_{4} / \mathcal{A}_{3}$ for the acceptance,
and \texttt{FCorr}~$\equiv$~$1-\eta-\frac{\eta}{\mathcal{M}}$ for the fragmentation.
Note that \texttt{ACorr} is the quantity shown in Fig.~\ref{Fig::HeliumAccRatios} (filled squares).
These values, to be applied as multiplicative factors to the fitted ratio, 
are also listed in Table~\ref{Tab::He3He4RatioSummary}.
Such corrections are affected by $\lesssim$3\% uncertainties in total, 
associated with the various physical and instrumental 
effects discussed here and in \S\ref{Sec::Acceptance}. 
All these errors and their role in the $^{3}$He/$^{4}$He ratio are reviewed in \S\ref{Sec::ErrorBreakdown}.

\subsubsection{$\delta$-Ray Emission} 
\label{Sec::DeltaRays}                

The effect of $\delta$-rays was included in our MC simulation.
In our previous work~\citep{AMS01Nuclei2010}, it was noted that energetic knock-on electrons 
affect the total acceptance at high energies. The production of $\delta$-rays is proportional
to the square of the primary particle charge, and the maximum energy of the produced
$\delta$-rays, $E^{\delta}_{\max}$, is proportional to the primary particle energy; for a nucleus of
momentum $M \gamma \beta c$, approximately: 
\begin{equation}
E^{\delta}_{\max} = 2 m_{e} c^{2} \beta^{2} \gamma^{2}
\label{Eq::DeltaRays}
\end{equation} 
For high energy nuclei ($E \gtrsim$ GeV~nucleon$^{-1}$), the emitted electrons can reach the 
anti-coincidence counters and veto the event, leading to an 
energy and charge dependent trigger efficiency. 
At lower energy, the $\delta$-rays curl up inside the tracking volume, 
affecting the reconstruction efficiency.
The influence of $\delta$-rays below 1.4 GeV~nucleon$^{-1}$ is negligible in this analysis and 
their effect has no significant difference between isotopes at the same energy, as $M \gg m_{e}$. 

\subsubsection{Background} 
\label{Sec::Background}    

The $Z=2$ charge separation from $Z<2$ and $Z>2$ samples 
was studied with MC simulations and
inflight data of e$^{-}$, p, He, and heavier ions. 
Proton (ion) beam data at CERN-PS (GSI)
provided additional validation at 0.75, 2.0, 3.6 and 8
GeV~nucleon$^{-1}$ of kinetic energy \citep{AMS01AntiHelium1999,AMS01Report2002,AMS01Nuclei2010}.
The main potential source of background to the helium sample 
was protons and deuterons wrongly reconstructed as $Z=2$ particles. 
Using the single TOF system or the single silicon tracker only, 
it can be seen, using flight data, that the probability of a $Z=1$
particle to be reconstructed as $Z=2$ is below 10$^{-3}$
thus affecting the helium sample of $\lesssim$1\%.
Using the combined measurements obtained from both the detectors,
the probability of the wrong charge magnitude was 
estimated to be $\sim\,$10$^{-7}$ 
over all energies \citep{AMS01AntiHelium1999}. 
Background from the less abundant $Z>2$ particles 
is completely negligible compared to the statistical 
uncertainties in the He sample (the He/Li ratio is $\sim\,$200 at the considered energies).

\subsubsection{Energy Losses and Resolution} 
\label{Sec::EnergyLosses}                    

Charged cosmic rays that traverse the detector lose energy in the
material above the tracking volume. The total energy loss has an
appreciable effect on the lowest energy bins.
The energy losses by $Z=2$ nuclei were estimated and parameterized 
with the MC simulation program. We made an event-by-event
correction to our data according to the average losses. 

Once the above corrections are performed, the relation between the reconstructed
energy of detected particles, $E${\tiny REC}, and their true energy, $E${\tiny TOI},
is still affected by the finite resolution of the measurement.
The probability of a bin-to-bin migration P($E${\tiny REC}$|E${\tiny TOI}) for He 
was estimated to effect only adjacent energy bins, to be symmetric and barely isotope dependent. 
Through these matrix elements, we estimated that the measured $^{3}$He/$^{4}$He ratio
has an uncertainty of 1--3\% due to the resolution.

\subsection{Uncertainty Estimate} 
\label{Sec::ErrorBreakdown}       

In Fig.~\ref{Fig::ccErrorBreakdownHelium}, we summarize the various sources 
of uncertainty in the measurement of the $^{3}$He/$^{4}$He. 
Errors are organized in four categories:  
\begin{figure}[!h]
\begin{center}
\epsscale{0.950}
\plotone{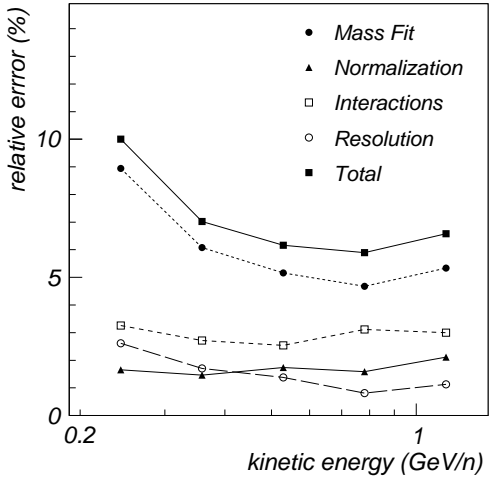}
\figcaption{ 
  Relative errors on the $^{3}$He/$^{4}$He ratio measurement as
  a function of the kinetic energy per nucleon.
  The total error (filled squares) is obtained by the sum in 
  quadrature of all the other contributions.
  The lines are to guide the eye.
  \label{Fig::ccErrorBreakdownHelium}
}
\end{center}
\end{figure}
\begin{deluxetable*}{l c ccccc c cccc}
  \centering
\tablecaption{
  Uncertainty summary for the measured isotopic ratios.
  The various contributions are described in \S\ref{Sec::ErrorBreakdown}.
  \label{Tab::ErrorBreakdownAll}
}
\tablehead{\\
  \colhead{Error} & \colhead{} &  \multicolumn{5}{c}{ $^{3}$He/$^{4}$He vs Energy (GeV/n) } &   
  \colhead{} & \multicolumn{4}{c}{ Ratios in 0.2\--1.4 GeV/n } \\
  \cline{1-1}\cline{3-7}\cline{9-12}\\
  \colhead{Type (Effect)} & \colhead{} & \colhead{0.2\--0.3}  & \colhead{0.4\--0.44} & \colhead{0.44\--0.64} & \colhead{0.64\--0.95} & \colhead{0.95\--1.4} & \colhead{} & \colhead{$^{3}$He/$^{4}$He} & \colhead{$^{6}$Li/$^{7}$Li} & \colhead{$^{7}$Be/$^{9+10}$Be} &\colhead{$^{10}$B/$^{11}$B}
}
\startdata 
Mass Fit ($\delta$A/A \& Statistics) &{}&  8.9~\% &  6.1~\% &  5.2~\% &  4.7~\% &  5.3~\%  &{}& 4.5~\% &  9.0~\% &  15.9~\% &  12.2~\% \\
Normalization ($\mathcal{N}$\--$\mathcal{N}_{\textrm{E}}$) &{}&   1.0~\% &  0.9~\%  &  1.3~\%  &  1.1~\% &  1.9~\% &{}& 1.5~\% &  2.3~\% &  3.0~\% &  2.9~\%  \\ 
Normalization (Acceptance) &{}&   1.3~\% &  1.2~\% &  1.1~\% &  1.1~\% &  1.0~\% &{}& 0.5~\% &  0.5~\% &  0.6~\% &  0.5~\% \\ 
Interactions (Inelastic) &{}&  1.6~\% & 1.6~\% & 1.8~\% & 1.9~\% & 2.0~\% &{}& 2.0~\% & 2.1~\% &  2.7~\% &  2.4~\% \\ 
Interactions (Fragmentation) &{}& 2.8~\% & 2.2~\%  &  1.8~\%  &  2.5~\%  &  2.3~\% &{}& 2.7~\%  &  2.4~\%  &  2.6~\%  &  2.4~\% \\ 
Resolution ($\delta \beta / \beta$) &{}& 2.6~\% &  1.7~\% &  1.4~\% &  0.8~\% &  1.1~\% &{}& 0.6~\% &  0.6~\% & 0.6~\% & 0.6~\% \\ 
Total Uncertainty &{}& 10.0~\% & 7.0~\% & 6.2~\% & 5.9~\% & 6.6~\% &{}& 5.8~\% & 9.8~\% & 16.7~\% & 13.0~\%  
\enddata
\end{deluxetable*}
\begin{enumerate}
\item \textit{Mass fit}. 
  The dominant source of uncertainty ($\sim\,$5$\--$9\%) is that associated to
  fits on the mass distributions (\S\ref{Sec::HeliumIsotopes}). 
  The errors were directly determined from the 1-$\sigma$ uncertainties 
  in the $\chi^{2}$ statistics of the fitting procedure. 
  These errors are due to statistical fluctuations of the measured data 
  and the inability of the spectrometer to separate the different masses 
  within the mass resolution $\delta$A/A. 
\item \textit{Normalization}. 
  As discussed in \S\ref{Sec::HeliumIsotopes}, two parameters $\mathcal{M}$ (ratio) 
  and $\mathcal{N}$ (normalization) were fitted. 
  An ideal fit should lead to $\mathcal{N}$ equal to the number of measured events, $\mathcal{N}_{\textrm{E}}$. 
  We took the relative difference $\mathcal{N}$\--$\mathcal{N}_{\textrm{E}}$ as a source of systematic error ($\sim\,$1\%). 
  Another contribution ($\sim\,$1\%) is due to the acceptance correction factors 
  estimated with our MC simulation program (\S\ref{Sec::Acceptance}). 
\item \textit{Interactions}. 
  Our results rely partially on hadronic interaction models, as discussed in \S\ref{Sec::NuclearInteractions}.
  Similarly to \citet{Wang2002}, we assumed an uncertainty of 10\% in the 
  inelastic cross sections,
  which corresponds to $\sim\,$2\% of systematic uncertainty.
  For the fragmentation channel $^{4}$He$\rightarrow$$^{3}$He,
  we assumed an uncertainty in the associated cross section equal to that cross section, 
  obtaining an uncertainty of 2$\--$3\% of the measured $^{3}$He/$^{4}$He ratio.
  Uncertainties in the material thickness were found to be negligible. 
\item \textit{Resolution}. 
  As discussed in \S\ref{Sec::EnergyLosses}, our measurement is affected by the finite 
  energy resolution of the TOF system. 
  A systematic uncertainty of 1$\--$3\% was estimated to account for this effect.
\end{enumerate}
The overall error (filled squares in Fig.~\ref{Fig::ccErrorBreakdownHelium}) is taken to be 
the sum in quadrature of the different contributions.
All these uncertainties are also reported in Table~\ref{Tab::ErrorBreakdownAll}.

\subsection{Lithium, Beryllium and Boron} 
\label{Sec::LiBeBIsotopes}                

In our previous work~\citep{AMS01Nuclei2010}, the lithium isotopic
composition was determined between 2.5 and 6.3 GV of magnetic rigidity.
Here we present a unified analysis of the lithium, beryllium and boron isotopes
between 0.2 and 1.4 GeV of kinetic energy per nucleon. 
In this measurement, we followed the same procedure as for the helium analysis.
All the steps described in \S\ref{Sec::HeliumIsotopes} were repeated for 
the study of the ratios $^{6}$Li/$^{7}$Li, $^{7}$Be/($^{9}$Be+$^{10}$Be) and $^{10}$B/$^{11}$B.
In this section we outline the essential parts of the $Z>2$ analysis.

The capability of the spectrometer to separate isotopes close in mass was more critical for $Z>2$
and the charge identification capabilities were limited by the 
use of tracker information only. 
However, the most limiting factor for the Li-Be-B study was the statistics. 
Hence, we performed the measurement with just one energy bin from 0.2 to 1.4 GeV~nucleon$^{-1}$, 
and also included data from  
the MIR-docking (b) and post-docking, non-nadir pointing (c) phases (\S\ref{Sec::AMS}). 
As in our previous work, for data collected during phase (b), a geometric cut 
on the MIR shadow was applied to the acceptance~\citep{AMS01Nuclei2010}. 
The geomagnetic regions considered and the event selection criteria 
were the same as for the helium analysis. 
Only four hits were required in the tracker, compared to five in the helium analysis.
The particle charge was assigned using the 
identification 
algorithm described in \citet{AMS01Nuclei2010} and \citet{Tomassetti2009},
that was specifically optimized for the $Z>2$ species.
\begin{figure*}[!ht]
\begin{center}
\epsscale{0.900}
\plotone{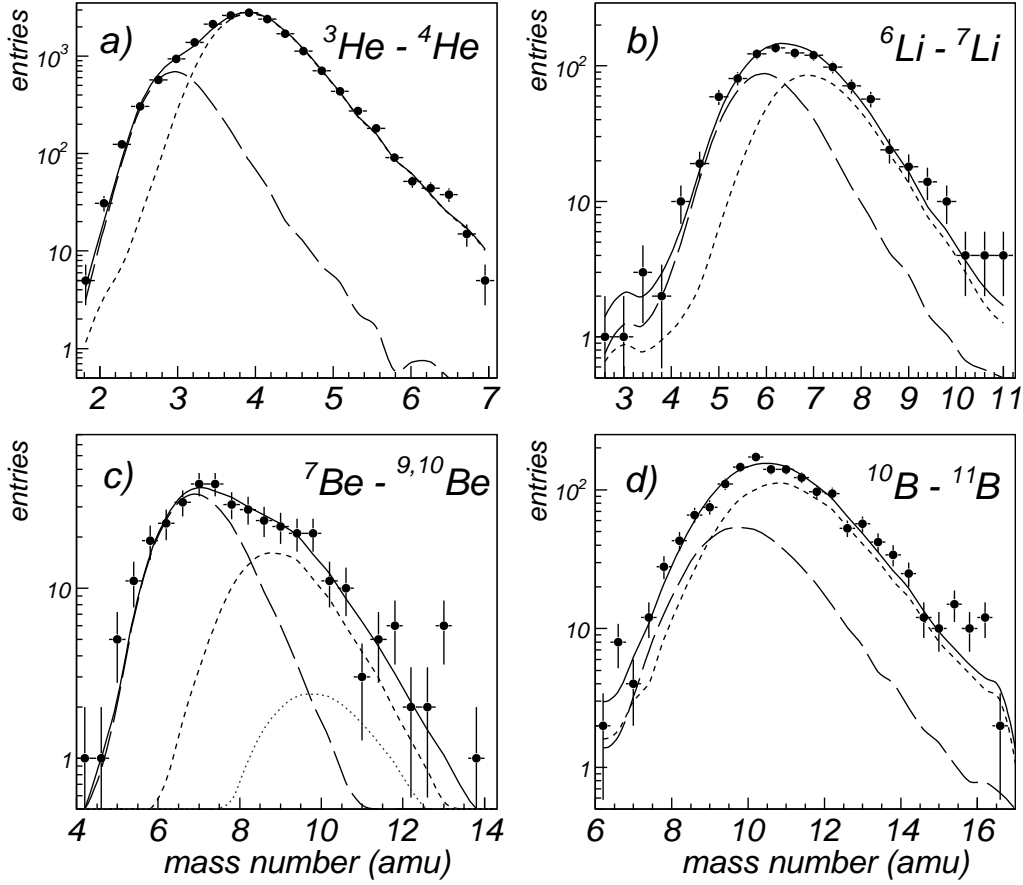}
\figcaption{ 
  Reconstructed mass distributions of events reconstructed as a) helium, 
  b) lithium, c) beryllium and d) boron from data (filled circles)
  and MC samples of $^{6}$Li$\--$$^{7}$Li, $^{7}$Be$\--$$^{9}$Be and $^{10}$B$\--$$^{11}$B 
  (lighter isotope: long-dashed lines; heavier isotope: short-dashed lines, sum: solid line).
  In c) the dotted line indicates a contribution of $^{10}$Be, see the text. 
  \label{Fig::ccRatioMassesLiBeB}
}
\end{center}
\end{figure*}

Results from the mass composition fit are shown in Fig.~\ref{Fig::ccRatioMassesLiBeB}. 
For comparison, the measurement was also performed on the average ratio of $^{3}$He/$^{4}$He over this energy range.
The large statistical fluctuations of the $Z>2$ data are apparent from the figure, 
in particular for the less abundant beryllium isotopes (400 events in total).
However, the agreement between the measured masses (filled circles) and the simulated 
histograms (solid lines) was satisfactory.
\begin{figure}[!h]
\begin{center}
\epsscale{0.950}
\plotone{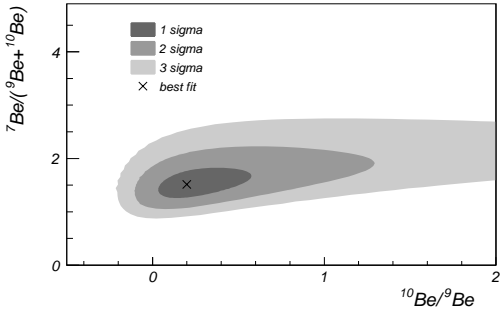}
\figcaption{ 
  Contour levels of the fitting parameter $^{7}$Be/($^{9}$Be+$^{9}$Be) and $^{10}$Be/$^{9}$Be 
  corresponding to 1\,$\sigma$, 2\,$\sigma$, and 3\,$\sigma$ bounds of uncertainty
  in the $\chi^{2}$ statistics.
  \label{Fig::ccPQCorrelation}
}
\end{center}
\end{figure}
The TOI corrections to the measured composition followed the procedure 
described in \S\ref{Sec::TOICorrections}.
In contrast to He, the $Z>2$ acceptances were found to be a bit smaller 
for the heavier isotopes ($^{7}$Li, $^{9,10}$Be and $^{11}$B) than the lighter ones ($^{6}$Li, $^{7}$Be and $^{10}$B),
indicating the dominance of nuclear interactions over other effects (see \S\ref{Sec::Acceptance}).
The $Z>2$ mass resolutions, $\delta$A/A, were found to be $\sim\,$8\% larger than for the $Z=2$ case,
reflecting the slight charge dependence of the spectrometer performance in particle tracking and timing. 
Corrections for fragmentation were also performed, considering the  
channels $^{7}$Li$\rightarrow$$^{6}$Li, $^{10,9}$Be$\rightarrow$$^{7}$Be and $^{11}$B$\rightarrow$$^{10}$B. 
\begin{deluxetable}{lrccccc}[!h]
\tablecaption{
  Fit results and correction factors for the ratios 
  $^{3}$He/$^{4}$He, $^{6}$Li/$^{7}$Li, $^{7}$Be/($^{9}$Be+$^{10}$Be) and $^{10}$B/$^{11}$B
  in the range $0.2\--1.4$ GeV of kinetic energy per nucleon. 
  \label{Tab::HeLiBeBRatioSummary}}
\tablehead{ 
  \colhead{Ratio} & \colhead{Events} & \colhead{Fit Results} & \colhead{$\chi^{2}$/df}  & 
  \colhead{$\delta$A/A} & \colhead{ACorr} & \colhead{FCorr}
}
\startdata
  $^{3}$He/$^{4}$He & 18,035 & 0.174 $\pm$ 0.009 & 67.7/34   & 12.7\% & 1.02 & 0.98 \\ 
  $^{6}$Li/$^{7}$Li & 1,046  & 0.951 $\pm$ 0.086 & 16.1/23   & 13.6\% & 0.97 & 0.99 \\
  $^{7}$Be/$^{9+10}$Be &  400  & 1.512 $\pm$ 0.238 & 19.9/23 & 13.8\% & 0.96 & 0.99 \\
  $^{10}$B/$^{11}$B & 1,598  & 0.494 $\pm$ 0.060 & 28.0/29   & 13.9\% & 0.96 & 0.99  
\enddata
\end{deluxetable}
A summary of fit results, TOI corrections and resolutions is given in Table~\ref{Tab::HeLiBeBRatioSummary}.

The errors were estimated as discussed in \S\ref{Sec::ErrorBreakdown}.  
The dominant error is that arising from the $\chi^{2}$ fit procedure: 
the small $Z>2$ statistics and the broader mass  
distributions led to less constrained composition fits.
Background from helium was estimated not to affect  
the lithium measurement, as for He$\--$Li charge separation the TOF 
information was still usable~\citep{AMS01Nuclei2010}.
More critical was the contamination in the $Z=4$ sample from adjacent
charges that led to larger systematic errors in the beryllium measurement.
This channel was also limited by the inability to separate $^{9}$Be from $^{10}$Be. 
While for the other ratios it can be safely assumed that each charged species is composed of 
only two long-lived isotopes, 
a few percent of $^{10}$Be has been measured in the cosmic ray flux
in addition to the more abundant isotopes $^{7}$Be and $^{9}$Be~\citep{Hams2004,Webber2002}. 
We therefore determined the ratio $^{7}$Be/($^{9}$Be+$^{10}$Be) 
simultaneously with the additional parameter $^{10}$Be/$^{9}$Be in our composition fit . 
The latter was accounted for the proper determination of 
the ratio $^{7}$Be/($^{9}$Be+$^{10}$Be) and its corresponding error.
As shown in Fig.~\ref{Fig::ccPQCorrelation}, the ratio $^{10}$Be/$^{9}$Be is poorly 
constrained by the data (between $\sim\,$0 and $\sim\,$0.6 within 1-$\sigma$ of uncertainty).
Fig.~\ref{Fig::ccPQCorrelation} also shows that the uncertainty in the $^{10}$Be/$^{9}$Be ratio
has no dramatic consequences in the $^{7}$Be/($^{9}$Be+$^{10}$Be) ratio, 
given the weak correlation of the two parameters.
The contribution from inelastic collisions and fragmentation was estimated as 
described in \S\ref{Sec::NuclearInteractions}.
Similar values as for helium ($\sim\,$2$\--$3\%) were found for both the effects.
Finally, 
the errors from the TOF energy resolution and 
from the MC acceptance estimation were smaller than 1\%.
The total error assigned to the measurements was obtained by the sum in
quadrature of all the noted contributions. 
A detailed summary is provided in Table~\ref{Tab::ErrorBreakdownAll}.

\section{Results and Discussion}  
\label{Sec::ResultsAndDiscussion} 

\begin{figure*}[!ht]
\begin{center}
\epsscale{0.900}
\plotone{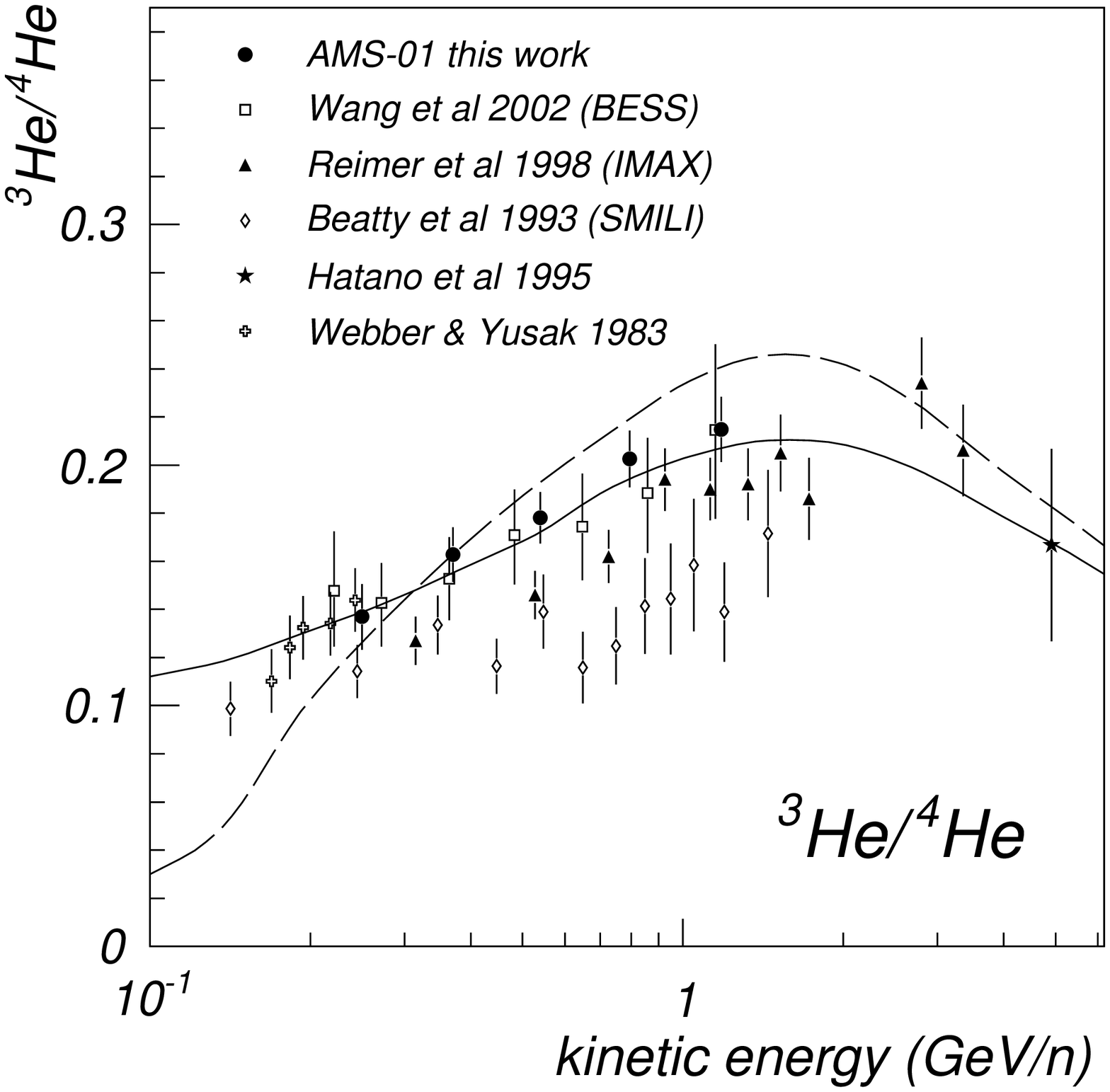}
\figcaption{ 
  Results for the ratio $^{3}$He/$^{4}$He   
  between 0.2 and 1.4 GeV~nucleon$^{-1}$ of kinetic energy per nucleon. 
  Other data are from BESS \citep{Wang2002}, IMAX \citep{Reimer1998}, \citet{Hatano1995}, 
  SMILI \citep{Beatty1993}, \citet{WebberYushak1983}.
  The dashed line is the model calculation for the LIS ratio obtained with \texttt{GALPROP}~\citep{GALPROP}.
  The modification of this by solar modulation for 1998 June is shown by the solid line.  
  \label{Fig::ccHeliumRatioResults}
}
\end{center}
\end{figure*}

In the previous sections, we have described the analysis procedure adopted
for the determination of the ratios $^{3}$He/$^{4}$He, $^{6}$Li/$^{7}$Li, $^{7}$Be/($^{9}$Be+$^{10}$Be) and $^{10}$B/$^{11}$B.
The TOI corrections turned out to be of the same order of magnitude as the estimated uncertainties,
hence, the gross features of the measured ratios were apparent directly from the fits on the mass distributions.
The error from the fitting procedure was considerably larger than the other contributions. 
The most important limitations were the mass resolution (for He) 
and the limited statistics (for Li-Be-B).
In particular, the mass resolution was limited by multiple scattering
(affecting $\delta R/R$ at $\sim\,$0.2 GeV~nucleon$^{-1}$) and the TOF
resolution (affecting $\delta \beta / \beta$ at $\sim\,$1.4 GeV~nucleon$^{-1}$).
\begin{deluxetable}{cc c cc}[!h]
\tablecolumns{5} 
\tablewidth{0pc} 
\tablecaption{
  Results for the isotopic ratios and fluxes
  at the top of instrument.
  \label{Tab::AllResults}} 
\tablehead{
  \colhead{Energy\tablenotemark{a}} & \colhead{$^{3}$He/$^{4}$He Ratio} & \colhead{}& \colhead{$^{3}$He Flux\,\tablenotemark{b}} & \colhead{$^{4}$He Flux\,\tablenotemark{b}}
}
\startdata
$0.20 \-- 0.30$\hfill &   0.137 $\pm$ 0.014 &{}& 23.3 $\pm$ 2.6 & 170 $\pm$  19    \\
$0.30 \-- 0.44$\hfill &   0.163 $\pm$ 0.011 &{}& 24.7 $\pm$ 2.2 & 152 $\pm$  14    \\
$0.44 \-- 0.64$\hfill &   0.178 $\pm$ 0.011 &{}& 21.6 $\pm$ 1.8 & 121 $\pm$  10    \\
$0.64 \-- 0.95$\hfill &   0.203 $\pm$ 0.012 &{}& 17.5 $\pm$ 1.4 & 86.5 $\pm$ 6.9  \\
$0.95 \-- 1.40$\hfill &   0.215 $\pm$ 0.014 &{}& 12.2 $\pm$ 1.0 & 56.6 $\pm$ 4.7  \\
\tableline \\
{ Energy\tablenotemark{a}} & { $^{2}$H/$^{4}$He Ratio} &{}& \multicolumn{2}{c}{ Ratios in 0.2$\--$1.4~GeV/n }\\
\cline{1-2}\cline{4-5}\\  
$0.20 \-- 0.30$\hfill & 0.183 $\pm$ 0.024 &{}& \multicolumn{1}{l}{$^{3}$He/$^{4}$He} & 0.173 $\pm$ 0.010 \\
$0.30 \-- 0.44$\hfill & 0.190 $\pm$ 0.020 &{}& \multicolumn{1}{l}{$^{6}$Li/$^{7}$Li} & 0.912 $\pm$ 0.090 \\
$0.44 \-- 0.64$\hfill & 0.188 $\pm$ 0.021 &{}& \multicolumn{1}{l}{$^{7}$Be/$^{9+10}$Be} & 1.450 $\pm$ 0.242 \\
$0.64 \-- 0.95$\hfill & 0.204 $\pm$ 0.027 &{}& \multicolumn{1}{l}{$^{10}$B/$^{11}$B} & 0.4695 $\pm$ 0.061   
\enddata
\tablenotetext{a}{Kinetic energy is given units of GeV~nucleon$^{-1}$.}
\tablenotetext{b}{Fluxes are given in units of nucleon/GeV/s/m$^{2}$/sr.}
\end{deluxetable}
The results with all corrections applied are presented in Table~\ref{Tab::AllResults}. 
Results for the isotopic ratio $^{3}$He/$^{4}$He as a function of the kinetic energy per nucleon
are shown in Fig.~\ref{Fig::ccHeliumRatioResults} between 0.2 and 1.4 GeV~nucleon$^{-1}$ (filled circles).
The error bars represent the total errors as discussed in \S\ref{Sec::ErrorBreakdown}.
The figure also shows the existing data between 0.1 and 10 GeV~nucleon$^{-1}$
measured by the balloon borne experiments 
BESS~\citep{Wang2002}, IMAX~\citep{Reimer1998}, the first flight of SMILI~\citep{Beatty1993}, 
\citet{Hatano1995} and \citet{WebberYushak1983}.
Among these, our data are the only data collected directly in space. 
Our results agree well with data collected by BESS in its first flight in 1993.

\begin{figure*}[!ht]
\begin{center}
\epsscale{0.900}
\plotone{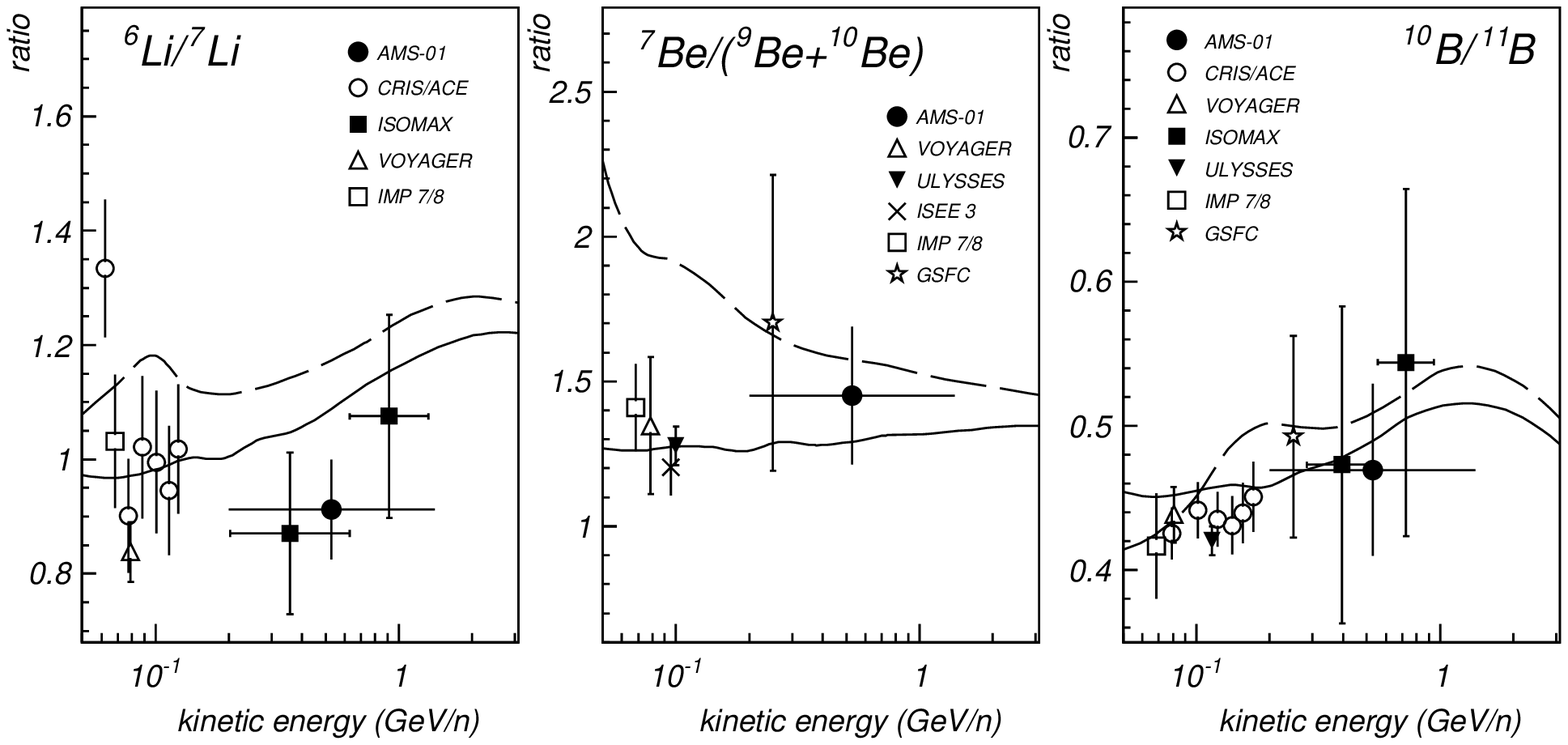} 
\figcaption{ 
  Results for the ratios $^{6}$Li/$^{7}$Li, $^{7}$Be/($^{9}$Be+$^{10}$Be) and $^{10}$B/$^{11}$B 
  between 0.2 and 1.4 GeV~nucleon$^{-1}$ of kinetic energy per nucleon
  ({\large $\bullet$}=AMS-01 present work, {\large $\circ$}=CRIS/ACE~\citep{DeNolfo2001}, 
  {\scriptsize $\blacksquare$}=ISOMAX~\citep{Hams2004}, $\vartriangle$=VOYAGER~\citep{Webber2002}, 
  $\blacktriangledown$=ULYSSES~\citep{Connell1998}, 
  $\times$=ISEE~3~\citep{Wiedenbeck1980},
  $\Box$=IMP~7/8~\citep{GarciaMunoz1977} and
  {\large$\star$}=GSFC~\citep{Hagen1977}).
  The dashed (solid) lines are the model calculations for the LIS (solar modulated) 
  ratios obtained with \texttt{GALPROP}~\citep{GALPROP}.  
  \label{Fig::ccIsotopicData3X1}
}
\end{center}  
\end{figure*}

Fig.~\ref{Fig::ccIsotopicData3X1} shows our results for Li-Be-B.
The AMS-01 data are compared with measurements made by the space experiments 
CRIS on ACE~\citep{DeNolfo2001}, VOYAGER~\citep{Webber2002}, 
ULYSSES~\citep{Connell1998}, ISEE~3~\citep{Wiedenbeck1980}, 
and with balloon data from 
ISOMAX~\citep{Hams2004},  
IMP~7/8~\citep{GarciaMunoz1977} and GSFC~\citep{Hagen1977}.
Our results are consistent with these data within uncertainties, in particular with 
ISOMAX.

\begin{figure*}[!ht]
\begin{center}
\epsscale{0.900}
\plotone{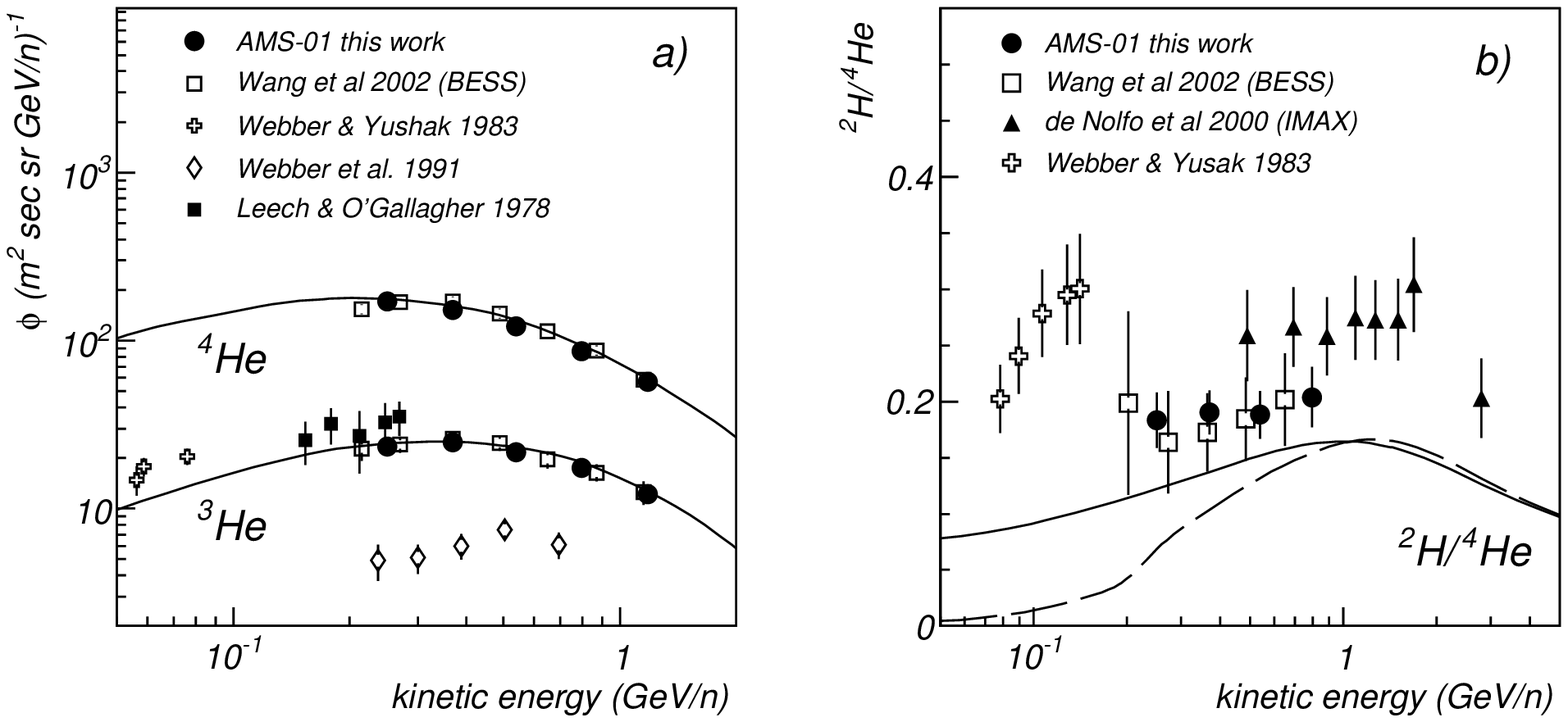}
\figcaption{ 
  a) Differential spectra of $^{3}$He and $^{4}$He. 
  Our data have been derived from our earlier published helium spectrum
  combined with the $^{3}$He/$^{4}$He ratio of this work. Errors are smaller than the size of the circles.
  b) The ratio $^{2}$H/$^{4}$He derived from our earlier work on $^{2}$H flux.
  In both figures, other data are from balloon borne experiments~\citep{Wang2002, WebberYushak1983, Webber1991, Leech1978,DeNolfo2000}.
  The dashed (solid) lines are the model calculations for the LIS (solar modulated) 
  spectra obtained with \texttt{GALPROP}~\citep{GALPROP}. 
  \label{Fig::ccCombinedResults}
}
\end{center}
\end{figure*}

In Fig.~\ref{Fig::ccCombinedResults}a we report the $^{3}$He and $^{4}$He
differential spectra. These spectra are obtained by the combination of 
the $^{3}$He/He and $^{4}$He/He fractions, directly given by the $^{3}$He/$^{4}$He ratio,
with the AMS-01 helium spectrum previously published in \citet{AMS01Helium2000}.
The $^{3}$He and $^{4}$He data points and their errors were extracted through a logarithmic interpolation 
of helium data along our energy points. 
An additional 1\% of error was added due to the interpolation procedure. 
Similarly, the resulting $^{4}$He spectrum has been further combined with the
galactic deuteron spectrum published in \citet{AMS01Report2002},
to extract the ratio between deuterons, $^{2}$H, and their main progenitors $^{4}$He. 
The AMS-01 data analysis of $^{2}$H is described in \S4.6 of \citet{AMS01Report2002},
where the extraction of the deuteron signal from the vast proton background 
is quantitatively discussed and the absolute deuteron spectrum is presented
in different geomagnetic latitude ranges.
The resulting $^{2}$H/$^{4}$He ratio is shown in Fig.~\ref{Fig::ccCombinedResults}b
together with the previous experiments BESS~\citep{Wang2002}, 
IMAX~\citep{DeNolfo2000}, and \citet{WebberYushak1983}.
While all the measurements give larger $^{2}$H/$^{4}$He ratios than the model 
predictions (see below) by up to a factor 2, our results again show good agreement with the data from BESS. 
These data are also reported in Table~\ref{Tab::AllResults}. 

To describe our data, we show in all plots the model calculations
of the conventional reacceleration model used in standard
methodologies and extensively described elsewhere~\citep{GALPROP}. 
Calculations have been made with the package \texttt{GALPROP-v50.1}\footnote[2]{
  \url{http://galprop.stanford.edu}
}.

\texttt{GALPROP} solves the diffusion\--transport equation for a given source distribution 
and boundary conditions for the galactic CRs, 
providing steady-state solutions for the local interstellar spectra (LIS) for all the charged CRs up to $Z=28$.
The diffusion of cosmic rays through the magnetic halo is described by means of a rigidity-dependent 
diffusion coefficient $D = \beta D_{0} \left( R/R_{0} \right)^{\delta}$, 
where $D_{0}$ and $R_{0}$ fix the normalization, and the spectral index $\delta$
drives its rigidity dependence.
The reacceleration of charged particles due to scattering on hydromagnetic waves 
is described as a diffusion in momentum space. 
This process is controlled by the Alfv\'en speed 
of plasma waves moving in the interstellar medium, $v_{A}$. 
The code also describes energy losses due to ionization or Coulomb scattering,
and catastrophic losses over the galactic disk, making use of a large compilation of 
cross section data and decay rates. To decouple all the transport equations,
the fragmentation network starts with the heaviest nucleus and works downward in mass,
processing primary and all secondary nuclei produced by the cascade. This loop is repeated twice.

In the parameter setting considered here, no tuning was done to our isotopic data. 
The nucleon injection spectrum is taken as a  ``broken'' power law in rigidity 
to better match our total helium and proton spectra from \citet{AMS01Report2002}.
Two indices $\nu_{1}$ and $\nu_{2}$ were used below and above $R_{B}$. 
The cross section database was extended using the updated cross section list 
from the version \texttt{v54} \citep{Vladimirov2011},
which includes the production of $^{2}$H and $^{3}$He 
from fragmentation of heavier isotopes.
The transport parameters $D_{0}$, $\delta$ and $v_{A}$ are consistent with our B/C ratio
from~\citet{AMS01Nuclei2010}. 
In our description, we used a 
cylindrically symmetric model of the galactic halo 
with radius $\mathcal{R}$=30 kpc and height $z_{h}$= 4 kpc.  
The relevant parameters are reported in Table~\ref{Tab::ModelParameters};
the remaining specifications are as in the file \texttt{galdef\_50p\_599278} provided with the package.

In the ratios of Fig.~\ref{Fig::ccHeliumRatioResults}, \ref{Fig::ccIsotopicData3X1} and \ref{Fig::ccCombinedResults},
local interstellar (dashed lines) and heliospheric propagated (solid lines) calculations are shown. 
The heliospheric modulation is treated using the force field approximation~\citep{Gleeson1968}, 
with $\phi=450\,$MV as the modulation parameter to characterize 
the modulation strength for 1998 June.  
This is also in accordance with the study performed in \citet{Wiedenbeck2009} over the full solar cycle 23. 
It should be noted, however, that the force field approximation has no predictive power, \textit{i.e.,}  
the value employed for the parameter $\phi$ is contextual to the propagation framework
adopted to predict the interstellar spectra of the CR elements.
For instance, the deuteron flux in \citet{AMS01Report2002} was described using $\phi=650\,$MV 
after assuming a pure power-law energy spectrum in the $^{2}$H LIS. 
\begin{deluxetable}{llc}[!h]
\tablecaption{
  Propagation parameter set.
  \label{Tab::ModelParameters}}
\tablehead{
  \colhead{Parameter} & \colhead{Name} & \colhead{Value}
}
\startdata
Injection, break value & $R_{B}$ [GV]     & 9 \\
Injection, index below $R_{B}$ &    $\nu_{1}$ & 1.80 \\  
Injection, index above $R_{B}$ &    $\nu_{2}$ & 2.35 \\

Diffusion, magnitude    &  $D_{0}$ [cm$^2$ s$^{-1}$]  & $5\cdot10^{28}$ \\
Diffusion, index        &   $\delta$                 & 0.41 \\
Diffusion, ref. rigidity & $R_{0}$ [GV]              & 4 \\

Reacceleration, Alfv\'en speed & $v_{A}$ [km s$^{-1}$]  & 32  \\

Galactic halo, radius  &   $\mathcal{R}$ [kpc] & 30  \\
Galactic halo, height  &   $z_{h}$ [kpc]        & 4   \\

Solar modulation parameter &  $\phi$ [MV]   & 450  
\enddata
\end{deluxetable}

Though large uncertainties are still present in the heliospheric propagation,
the general trend is that higher modulation levels correspond to lower values of 
the $^{3}$He/$^{4}$He ratio.
AMS-01 and BESS data come from periods of relatively quiet solar activity
as do data from \citeauthor{WebberYushak1983} ($\phi \approx 400\--650\,$MV). 
In particular, the periods of 1998 June (AMS-01 flight) and 1993 July (BESS flight) were
characterized by very similar solar conditions according to the sunspot data~\citep{Temmer2002}
and can be directly compared.
Stronger modulations were present when IMAX ($\phi \approx 700\--850\,$MV) 
and SMILI ($\phi \approx 1200\--1300\,$MV) were active. 

In summary, the secondary to primary ratio $^{3}$He/$^{4}$He, 
which is much more sensitive to the propagation parameters, 
seems to be well described by the model under these astrophysical assumptions. 
The model is also consistent with the $Z>2$ ratios of Fig.~\ref{Fig::ccIsotopicData3X1}, 
though large errors are present in the current data. 
As the Li-Be-B elements are of secondary origin, 
these ratios are less sensitive to the galactic transport, 
and may be useful to investigate the nuclear aspects of 
the CR propagation (fragmentation, decay, and breakup).
On the contrary, our $^{2}$H/$^{4}$He data of Fig.~\ref{Fig::ccCombinedResults}b 
give a larger ratio than the model predictions, 
and this tendency is also apparent from the other experiments.
Understanding this possible discrepancy may require a thorough investigation of the
$^{2}$H production cross sections,
in particular for the reactions induced by cosmic ray protons and helium nuclei. 

\section{Conclusions}    
\label{Sec::Conclusions} 

The AMS-01 detector measured charged cosmic rays during 10 days aboard the Space
Shuttle \textit{Discovery} in 1998 June. 
Owing to the large number of helium events collected and the absence of the atmospheric effects, 
we have precisely determined 
the ratio $^{3}$He/$^{4}$He in the kinetic energy range from 0.2 to 1.4 GeV\,nucleon$^{-1}$.
The average isotopic ratios $^{6}$Li/$^{7}$Li, $^{7}$Be/($^{9}$Be+$^{10}$Be) and $^{10}$B/$^{12}$B have
been measured in the same energy range. 
The ratio $^{2}$H/$^{4}$He and the spectra of $^{3}$He and $^{4}$He are also reported.
Our results agree well with the previous data from BESS and ISOMAX and can
provide further constraints to the astrophysical parameters of cosmic ray propagation.
In the analysis procedure adopted in this work,
Monte Carlo simulations were essential for understanding the instrument 
performance, its acceptance, the role of interactions and for modeling the mass distributions.
We expect, with AMS-02, to achieve much more precise results over wider energy ranges.

\section{Acknowledgements} 

The support of INFN, Italy, ETH-Zurich, the University of
Geneva, the Chinese Academy of Sciences, Academia Sinica
and National Central University, Taiwan, the RWTH Aachen,
Germany, the University of Turku, the University of Technology
of Helsinki, Finland, the US DOE and MIT, CIEMAT,
Spain, LIP, Portugal and IN2P3, France, is gratefully acknowledged.
The success of the first AMS mission is due to many individuals
and organizations outside of the collaboration. The support
of NASA was vital in the inception, development and operation
of the experiment. Support from the Max-Planck Institute for
Extraterrestrial Physics, from the space agencies of Germany
(DLR), Italy (ASI), France (CNES) and China and from CSIST,
Taiwan also played important roles in the success of AMS.
\\


\end{document}

%% file: authorlist_eapj.tex
\author{
M.~Aguilar$^{y}$, 
J.~Alcaraz$^{y}$,
J.~Allaby$^{r}$\altaffilmark{$\dagger$}, 
B.~Alpat$^{ad}$, 
G.~Ambrosi$^{ad}$, 
H.~Anderhub$^{aj}$,
L.~Ao$^{g}$, 
A.~Arefiev$^{ab}$, 
L.~Arruda$^{x}$,
P.~Azzarello$^{ad}$
% ...
}

\affil{\mysize $^a$ I. Physikalisches Institut, RWTH, D-52074 Aachen, Germany\altaffilmark{3}} 
\affil{\mysize $^b$ III. Physikalisches Institut, RWTH, D-52074 Aachen, Germany\altaffilmark{3}}   
\affil{\mysize $^c$ LAPP, Universit\'e de Savoie, CNRS/IN2P3, F-74941 Annecy-le-Vieux Cedex, France}
\affil{\mysize $^d$ Department of Physics and Astronomy, Johns Hopkins University, Baltimore, MD 21218, USA}
\affil{\mysize $^e$ Department of Physics and Astronomy, Louisiana State University, Baton Rouge, LA 70803, USA}
\affil{\mysize $^f$ Center of Space Science and Application, Chinese Academy of Sciences, 100080 Beijing, China}
\affil{\mysize $^g$ Chinese Academy of Launching Vehicle Technology, CALT, 100076 Beijing, China}
\affil{\mysize $^h$ Institute of Electrical Engineering, IEE, Chinese Academy of Sciences, 100080 Beijing, China}
\affil{\mysize $^i$ Institute of High Energy Physics, IHEP, Chinese Academy of Sciences, 100039 Beijing, China\altaffilmark{4}} 
\affil{\mysize $^j$ Dipartimento di Fisica and INFN, Universit\`a di Bologna, I-40126 Bologna, Italy\altaffilmark{5}}  
\affil{\mysize $^k$ Institute of Microtechnology, Politechnica University of Bucharest and University of Bucharest, R-76900 Bucharest, Romania}
\affil{\mysize $^l$ Massachusetts Institute of Technology, Cambridge, MA 02139, USA}
\affil{\mysize $^m$ National Central University, Chung-Li 32054, Taiwan }
\affil{\mysize $^n$ Institute for Physical Science and Technology, University of Maryland, College Park, MD 20742, USA}
\affil{\mysize $^o$ CHEP, Kyungpook National University, 702-701 Daegu, Republic of Korea}
\affil{\mysize $^p$ CNR--IROE, I-50125 Florence, Italy}
\affil{\mysize $^q$ Max--Planck Institut f\"ur extraterrestrische Physik, D-85740 Garching, Germany}
\affil{\mysize $^r$ European Laboratory for Particle Physics, CERN, CH-1211 Geneva 23, Switzerland}
\affil{\mysize $^s$ DPNC, Universit\'e de Gen\`eve, CH-1211 Geneva 4, Switzerland}
\affil{\mysize $^t$ LPSC, Universit\'e Joseph Fourier Grenoble 1, CNRS/IN2P3, Institut Polytechnique de Grenoble, 38026 Grenoble, France}
\affil{\mysize $^u$ Chinese University of Science and Technology, USTC, Hefei, Anhui 230 029, China\altaffilmark{4}} 
\affil{\mysize $^v$ Helsinki University of Technology, FIN-02540 Kylmala, Finland}
\affil{\mysize $^w$ Instituto Superior T\'ecnico, IST, P-1096 Lisboa, Portugal}   
\affil{\mysize $^x$ Laboratorio de Instrumentacao e Fisica Experimental de Particulas, LIP, P-1000 Lisboa, Portugal}
\affil{\mysize $^y$ Centro de Investigaciones Energ{\'e}ticas, Medioambientales y Tecnol\'ogicas, CIEMAT, E-28040 Madrid, Spain\altaffilmark{6}} 
\affil{\mysize $^{z}$ INFN-Sezione di Milano, I-20133 Milan, Italy\altaffilmark{5}} 
\affil{\mysize $^{aa}$ Kurchatov Institute, Moscow, 123182 Russia}
\affil{\mysize $^{ab}$ Institute of Theoretical and Experimental Physics, ITEP, Moscow, 117259 Russia}
\affil{\mysize $^{ac}$ Dipartimento di Fisica, Universit\`a Degli Studi di Perugia, I-06100 Perugia, Italy} 
\affil{\mysize $^{ad}$ INFN-Sezione di Perugia, I-06100 Perugia, Italy\altaffilmark{5}}  
\affil{\mysize $^{ae}$ Dipartimento di Fisica and INFN, Universit\`a di Pisa, I-56100 Pisa, Italy\altaffilmark{5}} 
\affil{\mysize $^{af}$ INFN-Sezione di Roma, I-00185 Roma, Italy\altaffilmark{5}}  
\affil{\mysize $^{ag}$ Department of Physics, Ewha Womens University, 120-750 Seoul, Republic of Korea}
\affil{\mysize $^{ah}$ Institute of Physics, Academia Sinica, Nankang Taipei 11529, Taiwan}
\affil{\mysize $^{ai}$ Space Research Laboratory, SRL, University of Turku, FIN-20014 Turku, Finland}
\affil{\mysize $^{aj}$ Eidgen\"ossische Technische Hochschule, ETH Z\"urich, CH-8093 Z\"urich, Switzerland}


\altaffiltext{$\bigstar$}{\mysize Corresponding author:~N.~Tomassetti (\href{mailto:Nicola.Tomassetti@pg.infn.it}{Nicola.Tomassetti@pg.infn.it})}

\altaffiltext{1}{\mysize Now at National Institute for High Energy Physics, NIKHEF, NL-1009 DB Amsterdam, The Netherlands.}

\altaffiltext{2}{\mysize Supported by ETH Z\"urich.}

\altaffiltext{3}{\mysize Supported by the 
Deutsches Zentrum f\"ur Luft-- und Raumfahrt, DLR.}

\altaffiltext{4}{\mysize Supported by the National Natural Science Foundation of China.}

\altaffiltext{5}{\mysize Also supported by the Italian Space Agency.}

\altaffiltext{6}{\mysize Also supported by the Comisi\'on Interministerial de Ciencia y Tecnolog{\'\i}a. }

\altaffiltext{$\dagger$}{\mysize Deceased.}